\documentclass[sigconf]{acmart}

\setcopyright{acmcopyright}

\usepackage{graphicx}
\usepackage{lscape}
\usepackage{upgreek}
\usepackage{caption} 
\usepackage{hyperref}
\usepackage{todonotes}
\usepackage{lineno}
\usepackage{url}
\usepackage{array}
\usepackage{enumitem}
\usepackage{multirow}
\usepackage{soul}
\usepackage{float}
\usepackage{color}
\usepackage{pgfplots}
\usepackage{booktabs}
\usepackage[export]{adjustbox}
\usepackage{colortbl}
\usepackage{fontawesome}
\usepackage[normalem]{ulem}

\acmDOI{xx.xxx/xxx_x}

\acmISBN{xxx-x-xxxx-xxxx-x/YY/MM}

\acmConference[XXX 2024]{XXX Conference}{Month 2024}{City, Country}
\acmYear{2024}
\copyrightyear{2024}

\acmArticle{4}
\acmPrice{15.00}

\begin{document}
\title{Modular Monolith: Is This the Trend in Software Architecture?}
  

\author{Ruoyu Su}
\affiliation{
\institution{M3S, University of Oulu}
   \city{Oulu} 
   \country{Finland}
 }
 \email{ruoyu.su@.oulu.fi}

 \author{Xiaozhou Li}
 \affiliation{
   \institution{M3S, University of Oulu}
  \city{Oulu} 
   \country{Finland}
 }
 \email{xiaozhou.li@oulu.fi}




\renewcommand{\shortauthors}{R. Su et al.}

\begin{abstract}
Recently modular monolith architecture has attracted the attention of practitioners, as Google proposed "Service Weaver" framework to enable developers to write applications as modular monolithic and deploy them as a set of microservices. Google considered it as a framework that has the best of both worlds and it seems to be a trend in software architecture. This paper aims to understand the definition of the modular monolith in industry and investigate frameworks and cases building modular monolith architecture. We conducted a systematic grey literature review, and the results show that modular monolith combines the advantages of monoliths with microservices. We found three frameworks and four cases of building modular monolith architecture. In general, the modular monolith is an alternative way to microservices, and it also could be a previous step before systems migrate to microservices.
\end{abstract}


\maketitle

\section{Introduction}


Microservices have been getting more and more popular in the last few years, and a large number of companies are migrating monolithic applications to microservices \cite{taibi2017processes}. Especially in the industry, companies like Amazon and Netflix are looking to take advantage of microservices such as independent development and deployment to help their systems solve problems \cite{lenarduzzi2020does}. These problems are mostly due to the drawbacks of legacy and monolithic systems that the direct tight coupling of their internal components makes them difficult to maintain and increases the development time and effort required \cite{taibi2019monolithic}.

However, several companies did not get the expected benefits from migrating to microservices, and fell into difficulty because of issues such as high cost and complexity of microservices \cite{su2023back}. Since AWS Re-Invent 2018 started to mention that the migration to microservices blindly was a mistake \cite{online1}, more and more practitioners have started discussing the topics that migrating from monolith to microservices. Even there are examples of switching from microservices back to the monolith like Amazon PrimeVideo \cite{online2}.In general, monoliths and microservices are not "perfect" and they have their own different drawbacks \cite{su2023back}.

Recently, the concept of "Modular Monolith" has attracted the attention of practitioners, as Google proposed the "Service Weaver" framework to enable developers to write applications as modular monolithic and deploy them as a set of microservices \cite{online3}. Google explains that it is a framework that has the best of both worlds: the development velocity of a monolith, with the scalability, security, and fault-tolerance of microservices \cite{online3}.

Traditional monoliths focus on layers and often include three layers: UI, Business and Data \cite{online4}. All features are vertically separated into these layers. The business layer is the one that contains the business logic of all features. Each feature knows the business logic of other features, which are tightly coupled \cite{online4}.

The concept of modularization is not new knowledge, it has been proposed in the last century that modularization is a mechanism for improving the flexibility and comprehensibility of systems \cite{parnas1972criteria}. It differs from a monolithic system in that modularization divides the system into separate modules, and independent teams can work on each module so that it reduces product development time and has greater flexibility and comprehensibility \cite{parnas1972criteria}.

The proposal of modular monolith is exciting, it has similarities with the traditional monolith and modularization mechanism proposed in the past but is different. It seems to be the "middle ground", the combination of monolith and microservices that addresses the issues of monoliths and microservices. The recently proposed Service Weaver framework has triggered even more discussion among practitioners about modular monolithic architectures.

This paper aims to understand the definition of the modular monolith in industry and investigate frameworks and cases in building modular monolith architecture. Based on our goals, we define the following research questions: \textit{\textbf{RQ$_1$} What is Modular Monolith? \textbf{RQ$_2$} Are there any frameworks proposed in building modular monolith architecture? \textbf{RQ$_3$} Are there any cases that use modular monolith architecture?} We conducted a systematic grey literature review and finally 64 results were included.

The results show that Modular Monolith is a software architecture pattern that combines the advantages of monolith with microservices architecture. In this architecture, systems are organized into loosely coupled modules, each delineating well-defined boundaries and explicit dependencies on other modules. Especially, it differs from the previously mentioned modularity in that it can be moved or deployed as microservices later if want. We found three frameworks for building the modular monolith: Service Weaver, Spring Modulith, and Light-hybrid-4j. There are also four cases that use modular monolith rather than microservices: Shopify, Appsmith, Gusto(Time Tracking), and PlayTech(Casino Backend Team). However, they do not use the frameworks proposed above. Through these results, we conclude that modular monolith is an alternative way to microservices, and it also could be a previous step before systems migrate to microservices.

The remainder of this paper is organized as follows: Section 2 reports the research method employed to conduct a systematic grey literature review. In Section 3, we analyze the results that address the goal of the study. Finally, Section 4 concludes the paper and outlines our future work.

\section{Methodology}

In this paper, we aim to understand the definition of the modular monolith in industry and investigate frameworks and cases in building modular monolith architecture. To such an end, we conducted a systematic grey literature review (SGLR) based on the guidelines defined by \cite{garousi2019guidelines}\cite{keele2007guidelines}. Herein, we used the following query in the grey literature search based on the identified keywords:

\begin{center}
    \emph{(serviceweaver OR ``service weaver'' OR ``modular monolith'' OR ``modular monolithic'')}
    \\\emph{AND}
    \\\emph{(``microservice'' OR ``microservices'')}
\end{center}

By searching on Google, we obtained 140 results in total. Each result was read and evaluated by two authors and the inclusion criteria is that the studies mention definitions, frameworks or cases on modular monolith. In the case of disagreement, both authors would discuss together and decide whether to include or exclude these results. We also adopted the "snowballing" method in the process to extract more relevant information \cite{wohlin2014guidelines}.

\section{Results}

The following section presents the results of the systematic grey literature review. According to the grey literature review, 64 studies are finally included that mention the definition, frameworks and cases on modular monolith. The included 64 studies(Ss) can be found in APPENDIX A.

\begin{figure}[h]
    \centering
    \includegraphics[width=\linewidth]{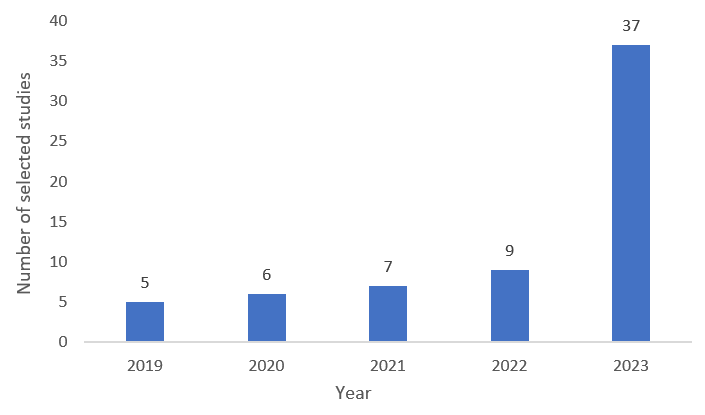}
    \caption{Selected Studies By Year}
    \label{fig:1}
\end{figure}

When dividing these selected 64 studies by the publication year, we can observe the trends of the discussion on modular monolith. As shown in Fig.\ref{fig:1}, the discussion about the topic of modular monolith started in 2019, and the number of selected studies is stable until 2022. However, its number increases dramatically in 2023. This trend indicates that modular monoliths have become a hot topic and have sparked intense discussion among practitioners. 

\subsection{RQ$_1$ What is Modular Monolith?}
Of the included 64 results, 47 studies answered the RQ1. We extracted data with four different parts of the modular monolith: Concept, Characteristics and features, Code structure and Testing.

\begin{itemize}[topsep=1pt, partopsep=0pt, itemsep=0pt, parsep=0pt]
    \item \textbf{Concept}
\end{itemize}

Modular Monolith is a software architecture pattern that strategically combines the simplicity of a monolithic structure with the advantages of microservices. In this approach, the system is organized into loosely coupled modules, each delineating well-defined boundaries and explicit dependencies on other modules. The goal is to achieve independence and isolation for each module, allowing them to be worked on independently while still being deployed collectively as a single unit. This architectural style seeks to strike a middle ground between traditional monolithic and microservices architectures. It emphasizes the interchangeability and potential reusability of modules, promoting a clear programming interface between them. The focus on business domains rather than technical layers, along with the vertical stacking of modules, enhances code organization and maintainability. Ultimately, a Modular Monolith represents a holistic and flexible approach to application design, where each module encapsulates specific functionality, fostering ease of development, testing, and deployment. Crucially, it can continue to be migrated to microservices or remain unchanged, which is more convenient than migrating directly from monolith to microservice [S1-S44].

\begin{itemize}[topsep=1pt, partopsep=0pt, itemsep=0pt, parsep=0pt]
    \item \textbf{Characteristics and features}
\end{itemize}

We found 6 studies that mention the characteristics, and we summarized 6 characteristics of the modular monolith:
(1) Segregation of modules. Each module is independent with its own layers such as Domain, Infrastructure and API. Modules are autonomously developed, tested, and deployed, affording the flexibility to employ diverse database solutions \ref{S18}\ref{S24}.
(2) Modularity with loose coupling and high cohesion. Modules exhibit loose interdependence and strong internal cohesion. Communication between modules occurs through APIs, preferably adopting loosely coupled asynchronous communication patterns \ref{S1}\ref{S18}\ref{S24}\ref{S46}.
(3) Unified Database Schema. The system adheres to a singular database schema, in contrast to microservices, where each microservice necessitates an individual schema \ref{S1}.
(4) Monolithic Deployment Structure. All modules within the modular monolith operate within the same Virtual Machine (VM), or each module may run on dedicated VMs. The scale of modules renders them impractical to be encapsulated within containers \ref{S1}.
(5) Unified Application Process. The application functions as a singular process, offering a uniform solution applicable to diverse scales of applications. Notably, there exists no rigid data ownership delineation among modules \ref{S22}\ref{S24}\ref{S26}.
(6) Enhanced Maintainability and Scalability. Comparative to traditional monolithic architectures, the modular monolith model demonstrably enhances both maintainability and scalability, underscoring its efficacy in managing complexities and facilitating growth \ref{S24}.

We also found 9 features of modular monolith in the selected studies \ref{S4}\ref{S5}\ref{S11}\ref{S17}\ref{S62}: 
(1) A module is never completely independent. It has dependencies with other modules, but these should be minimal; 
(2) Modules are interchangeable. 
(3) Code is reusable. 
(4) Better organization of the dependencies compared to traditional monolithic apps. 
(5) Easier to maintain and develop new versions than traditional monolithic apps. 
(6) You can keep the whole project as a single unit, without needing different servers for deployment. 
(7) More scalable than traditional monolithic apps. 
(8) Less complex than microservices architecture.
(9) Have defined API, allowing access to the logic of each module through public methods, not internal functions and the logic of each one.

\begin{itemize}[topsep=1pt, partopsep=0pt, itemsep=0pt, parsep=0pt]
    \item \textbf{Code structure}
\end{itemize}

In modular monolith architecture, the code should contain multiple functional modules, each module having an interface that represents its public definition \ref{S10}. In the selected studies \ref{S10}\ref{S12} \ref{S13}, authors proposed the general modular monolith architecture structure that exposes module interfaces in two ways: Externally, the module offers an API via REST HTTP or GRPC, with API calls managed by a proxy or gateway. Internally, services access the module through an abstracted interface, enabling information retrieval without direct access to the implementation. This upholds a clear separation of concerns, preserving application processes.


\begin{itemize}[topsep=1pt, partopsep=0pt, itemsep=0pt, parsep=0pt]
    \item \textbf{Testing}
\end{itemize}

Different from traditional monolithic systems but similar to the microservices systems, the modular monolith should have its own set of unit testing for each module to ensure that its functionality works as expected in isolation because it divides the system into modules \ref{S14}. The interaction between each module should also be verified by performing integration testing. In addition, developers also need to review and refactor the modular monolith on a regular basis to keep the code base neat maintainable and adaptable to changing business requirements \ref{S14}.

\subsection{RQ$_2$ Frameworks proposed in building modular monolith architecture}

Of the included 64 results, 17 studies answered the RQ2, and there are three frameworks proposed for building modular monolith architecture: Service Weaver, Spring Modulith, and Light-hybrid-4j.

\begin{itemize}[topsep=1pt, partopsep=0pt, itemsep=0pt, parsep=0pt]
    \item \textbf{Service Weaver}
\end{itemize}

Service Weaver is an open-source framework, written by Google, for building and deploying distributed systems. It provides the idea of decoupling the code from how code is deployed \ref{S45}\ref{S51}\ref{S52} \ref{S53}\ref{S54}\ref{S55}. The framework is currently only available in Go. It allows people to write the application as a modular monolith and deploy it as a set of microservices \ref{S58}\ref{S59}\ref{S60}\ref{S62}\ref{S63}\ref{S64}.

Service Weaver consists of two core pieces: A set of programming libraries that allow developers to write their application as a single modular binary using only native data structures and method calls; A set of deployers that allow developers to configure the runtime topology of their application and deploy it as a set of microservices, either locally or on the cloud of their choice \ref{S51}\ref{S54}\ref{S59}\ref{S60}\ref{S64}.

The motivation for building Service Weaver is Google found that the expense of maintaining multiple different microservice binaries with separate configuration files, network endpoints, and serializable data formats significantly slowed down the development of microservice-based applications \ref{S51}\ref{S59}\ref{S60}. Furthermore, microservices heightened the difficulty of cross-binary changes and rendered API modifications very challenging \ref{S54}\ref{S59}\ref{S60}. Therefore, developers hoped the monolithic binaries would be easy to write, update, and run locally or in a VM \ref{S54}\ref{S63}\ref{S64}.

The goal of Service Weaver is to improve distributed application development velocity and performance \ref{S59}\ref{S60}\ref{S63}\ref{S64}. Its core idea is a modular monolith model that creates a unified binary using native language data structures and methods, organized into modules (components) implemented as native types \ref{S51}\ref{S63}\ref{S64}.

Service Weaver has the following steps: First is to split the application into components written as regular Go interfaces. Secondly, call the components using regular Go method calls and no need for RPCs or HTTP requests. Then, test the application and deploy it to the cloud. Finally, run the components wherever in the same or different process and scale up or down to match load \ref{S52}\ref{S54}\ref{S64}.

Service Weaver has more powerful features compared to traditional modular monoliths. Service Weaver has highly performant in that co-located components communicate via direct method call while
remote components communicate using highly efficient custom serialization and RPC protocols \ref{S52}\ref{S62}. It has minimal configuration and deploys to the cloud without extensive boilerplate configuration \ref{S52}. Service Weaver provides libraries logging, metrics, and tracing, automatically integrated into the deployed cloud environment \ref{S52}\ref{S53}. In addition, it can do the effective sharding that shard requests across various component replicas for optimized performance \ref{S52}\ref{S62}. More importantly, Service Weaver has flexible scalability, it can easily scale applications horizontally or vertically on demand, whether as a monolithic application or a distributed microservice \ref{S52}.

However, the Service Weaver framework is still in development and now only has the v0.1 release which includes: The core Go libraries used for writing your applications; A number of deployers used for running your applications locally or on GKE; A set of APIs that allow you to write your own deployers for any other platform \ref{S51}\ref{S54}\ref{S59}\ref{S60}\ref{S64}.

\begin{itemize}[topsep=1pt, partopsep=0pt, itemsep=0pt, parsep=0pt]
    \item \textbf{Spring Modulith}
\end{itemize}

Spring Modulith is an experimental Spring project designed for modular monolith applications, with its source code organized based on the module concept \ref{S38}\ref{S43}\ref{S44}\ref{S48}. It provides conventions and APIs for declaring and validating logical modules within the Spring Boot application \ref{S44}. 

Spring Modulith ensures a modular structure for Spring Beans to grant control over what to expose. The core concept revolves around application modules, representing units of functionality with exposed APIs. Expressing modules can be done through various methods, including organizing domain or business modules as direct sub-packages. Module encapsulation is a key feature, safeguarding internal implementations through sub-packages of the application module's base package. Additionally, it restricts type visibility between modules, allowing access to module content \ref{S38}\ref{S43}\ref{S44}.

The first release of Spring Modulith introduces advanced features, including enhanced package arrangements, flexible module selection for integration tests, a transaction event publication log for seamless integration, automatic developer documentation generation with diagrams, runtime observability at the module level, and Passage of Time Events implementation \ref{S44}.

\begin{itemize}[topsep=1pt, partopsep=0pt, itemsep=0pt, parsep=0pt]
    \item \textbf{Light-hybrid-4j}
\end{itemize}

The light-hybrid-4j framework is designed to be a modular monolith framework, built on top of the light-4j for modularized monolithic and serverless architecture from the Light platform \ref{S37}. The benefit of using this framework when building a modular monolith system is gives flexibility for deployment and saves production costs \ref{S37}.

The framework involves creating a server with integrated third-party dependencies and building multiple services with shared or specific dependencies. Once services are developed, they are compiled into independent jar files containing only business handlers. These jars are placed in a designated folder on the host running the server, and upon server startup, all services in the folder are loaded. Traffic is then routed to the appropriate handler in the respective service for incoming requests. Developers adhere to the principle of building services into jar files, deploying them to the same Docker container volume, and loading them into the same JVM. Services communicate through interfaces, concealing implementation details. If needed, services experiencing heavy loads can be separated into individual containers for scalable deployment \ref{S37}.

\subsection{RQ$_3$ Cases that use modular monolith architecture}

Of the selected 64 studies, 6 studies answered the RQ3, and there are four cases use modular monolith architecture rather than microservices(specifically emphasized this): Shopify, Appsmith, Gusto (Time Tracking) and PlayTech (Casino Backend Team).

\textbf{Shopify} is one of the largest Ruby on Rails codebases in existence, worked on for over a decade by more than a thousand developers \ref{S12}. It was initially built as a monolith all the distinct functionalities were built into the same codebase with no boundaries between them. To better development, Shopify chose to change its architecture. Shopify originally considered microservices but finally chose modular monolith to balance the benefits of a single codebase with clear component boundaries. Therefore, Shopify is a great example that has used this approach as an alternative to microservice decomposition \ref{S12}\ref{S30}.

\textbf{Appsmith} is a platform that helps developers build internal apps \ref{S34}.They originally considered a microservice-based architecture, However, they rejected this alternative due to many considerations for on-premise deployment of a microservice-based application. Given their core on-premise deployment as a focal point of their business, they deemed it imperative to avoid introducing complexity into the deployment process for end users \ref{S33}\ref{S34}. Therefore, they chose modular monolith because it allowed them to maintain one large codebase, deploy a single binary, and balance the interests of their end customers with those of our internal team and open-source community \ref{S33}\ref{S34}.

\textbf{Gusto} wanted to build a new feature named "TimeTracking" and it would be a separate domain that has its own service \ref{S39}. To achieve this feature, they chose modular monolith, because for Gusto, their most important is to have a clear API boundary and make sure the system should not pass rich objects across that boundary \ref{S39}. They did not choose microservices for it has their own sets of challenges in testing and deploying and if got the service boundary wrong, it’d be much more expensive to fix \ref{S39}.

\textbf{PlayTech (Casino Backend Team)} discovered their company has many departments running microservices and cloud-native applications in the process of continuous refactoring. However, from the perspective of a single team, this maintenance would be unnecessary overhead. Therefore, their team chose a modular monolith as a result \ref{S32}.

In general, Shopify, Appsmith, Gusto (Time Tracking) and PlayTech (Casino Backend Team) are great examples that have used modular monolith as an alternative to microservice decomposition. However, they do not use the frameworks proposed in RQ2 above.

\section{Conclusion}

The modular monolith architecture has attracted the attention of practitioners for the "Service Weaver" framework proposed by Google. It seems to be a framework that has the best of both worlds: the development velocity of a monolith, with the scalability, security, and fault-tolerance of microservices\cite{online3}. We conducted a systematic grey literature review, which aims to understand the definition of the modular monolith in industry and investigate frameworks and cases in building modular monolith architecture. Starting from 140 results, 64 related studies were selected. The results show that modular monolith is a software architecture pattern that combines the advantages of monolith with microservices architecture and in particular, it can be moved or deployed as microservices later if want. In this architecture, systems are organized into loosely coupled modules, each delineating well-defined boundaries and explicit dependencies on other modules. We found three frameworks: Service Weaver, Spring Modulith and Light-hybrid-4j, and four cases: Shopify, Appsmith, Gusto(TimeTracking) and PlayTech(Casino Backend Team) in building the modular monolith architecture. We conclude that modular monolith is an alternative way to microservices, and it also could be a previous step before systems migrate to microservices. The study helps researchers and practitioners to deeply explore modular monoliths, and provides guidance in software architecture decisions, offering useful directions for future work. In our future work, we shall continue our research on the relationship and migration between modular monoliths and microservices to help organizations enable more effective implementation of architectural changes.





\bibliographystyle{ACM-Reference-Format}
\bibliography{bibliography}


\appendix
\section*{Appendix A: The Selected studies} 
\label{The Selected Papers}

{\small
 \begin{enumerate} [labelindent=-5pt,label={[S}{\arabic*]}]

\item \label{S1}
Itiel Maayan. Will Modular Monolith Replace Microservices Architecture? 2022. https://medium.com/att-israel/will-modular-monolith-replace-microservices-architecture-a8356674e2ea

\item \label{S2}
Mehmet Ozkaya. Microservices Killer: Modular Monolithic Architecture. 2023. https://medium.com/design-microservices-architecture-with-patterns/microservices-killer-modular-monolithic-architecture-ac83814f6862

\item \label{S3}
Rahul Garg. When (‌modular) monolith is the better way to build software. 2023. https://www.thoughtworks.com/insights/blog/micro-services/modular-monolith-better-way-build-software

\item \label{S4}
Sergio Agamez. Modular Monolithic vs. Microservices. https://www.-fullstacklabs.co/blog/modular-monolithic-vs-microservices

\item \label{S5}
Roshan Manjushree Adikari. Is modular monolith enough for your organization instead of microservices? 2023. https://levelup.gitconne-cted.com/is-modular-monolith-enough-for-your-organization-inste-ad-of-microservices-97917fa07f27

\item \label{S6}
Alex Bolboaca. Modular Monolith Or Microservices? 2019. https://m ozaicworks.com/blog/modular-monolith-microservices

\item \label{S7}
Adrian Kodja. Modular Monoliths vs. Microservices. 2023. https://ad riankodja.com/modular-monoliths-vs-microservices

\item \label{S8}
Adrain Ababei. Modular Monolith vs Microservices: How Do You Make a Choice? 2023. https://www.cmsdrupal.com/blog/modular-monolith-vs-microservices-how-do-you-make-choice

\item \label{S9}
Priyank Gupta. Understanding the modular monolith and its ideal use cases. 2020. https://www.techtarget.com/searchapparchitecture/ tip/Understanding-the-modular-monolith-and-its-ideal-use-cases

\item \label{S10}
Miłosz Lenczewski. What is better? Modular Monolith vs. Microservices. 2022. https://careers.tidio.com/blog-post/what-is-better-modul ar-monolith-vs-microservices

\item \label{S11}
Kirsten Westeinde. Deconstructing the Monolith: Designing Software that Maximizes Developer Productivity. 2019. https://shopify.en gineering/deconstructing-monolith-designing-software-maximizes-developer-productivity

\item \label{S12}
Chris Richardson. How modular can your monolith go? Part 1 - the basics. 2023. https://microservices.io/post/architecture/2023/07/31/ how-modular-can-your-monolith-go-part-1.html

\item \label{S13}
Dennis John. Modular Monolith - primer. 2023. https://www.linkedin. com/pulse/modular-monolith-primer-dennis-john/

\item \label{S14}
JRebel. What Is a Modular Monolith? 2020. https://www.jrebel.com/ blog/what-is-a-modular-monolith

\item \label{S15}
Drawing Boxes. Microservices vs Monolithic Architecture. 2023. https://www.youtube.com/watch?v=6-Wu178sOEE

\item \label{S16}
Matthew Freshwaters. Why Companies are Moving their Applications to Modular Architecture. 2023. https://blog.ippon.tech/why-companies-are-moving-their-applications-to-modular-architecture/

\item \label{S17}
Ahmet Kucukoglu. What is Modular Monolith? 2022. https://www. ahmetkucukoglu.com/en/what-is-modular-monolith

\item \label{S18}
Tomas Fernandez. 12 Ways to Improve Your Monolith Before Transitioning to Microservices. 2023. https://semaphoreci.com/blog/mono-lith-microservices

\item \label{S19}
KAMIL GRZYBEK. MODULAR MONOLITH: A PRIMER. 2019. https: //www.kamilgrzybek.com/blog/posts/modular-monolith-primer

\item \label{S20}
Debug Agent. Its Time to go Back to the Monoliths. Use Modular Monolith, save costs. 2023. https://www.youtube.com/watch?v= NWu7AJJlLM8

\item \label{S21}
Fernando Doglio. Domain-Driven Design for a Modular Monolith: Bridging the Gap Between Microservices and Monoliths. 2023. https://javascript.plainenglish.io/domain-driven-design-for-a-mod-ular-monolith-bridging-the-gap-between-microservices-and-monol-iths-2d2521196dd8

\item \label{S22}
Jinal Desai. Monolith vs Modular Monolith vs SOA vs Microservices vs MUI vs Nano Services. 2023. https://jinaldesai.com/monolith-vs-modular-monolith-vs-soa-vs-microservices-vs-mui-vs-nano-services/

\item \label{S23}
KnowledgeZone Admin. What is Modular Monolith? 2023. https:// knowledgezone.co.in/posts/632d68ee8a9e14d7a6c36e40

\item \label{S24}
Md Kamaruzzaman. Looking Beyond the Hype: Is Modular Monolithic Software Architecture Really Dead? 2020. https://towardsdata-science.com/looking-beyond-the-hype-is-modular-monolithic-soft-ware-architecture-really-dead-e386191610f8

\item \label{S25}
My experience of using modular monolith and DDD architectures. 2021. https://www.thereformedprogrammer.net/my-experience-of-using-modular-monolith-and-ddd-architectures/

\item \label{S26}
Application Developer from ThoughtWorks. Modular Monolith - A Step Towards Microservice. 2019. https://akshantalpm.github.io/Mo-dular-Monolith/

\item \label{S27}
Dzmitry Chubryk. Developing Modular Monolith vs. Traditional Monolith in Software Engineering, pros and cons. 2023. https://dev.to /flashblack/developing-modular-monolith-vs-traditional-monolith-in-software-engineering-pros-and-cons-p61

\item \label{S28}
Hacker News. https://news.ycombinator.com/item?id=29578712

\item \label{S29}
Arpit Mohan. Our Journey from SaaS to OSS: Embracing the Modular Monolith over Microservices. 2023. https://www.appsmith.com/blog/ monoliths-vs-microservices-1

\item \label{S30}
Thinktecture Team. Modular Monoliths With ASP.NET Core – Pragmatic Architecture. 2021. https://www.thinktecture.com/en/asp-net-core/modular-monolith/

\item \label{S31}
ThanhSangLuskillaug. Modular Monolithic (according to microservice's principles). 2022. https://www.slideshare.net/ThanhSangLusk-illaug/modular-monolithic-according-to-microservices-principles

\item \label{S32}
Light Platform. Modular Monolith. 2021. https://www.networknt. com/architecture/modular-monolith/

\item \label{S33}
Jakub Pierzchlewicz. THE MYTHICAL MODULAR MONOLITH. 2020. https://jpintelli.com/2020/09/03/the-mythical-modular-mono-lith/

\item \label{S34}
Maciej “MJ” Jedrzejewski. Dillema: Modular monolith or microservices? 2022. https://meaboutsoftware.com/2022/07/18/dillema-1-mo-nolith-or-microservices/

\item \label{S35}
Sebastian. Modular monolith in Python. 2020. https://breadcrumbsc-ollector.tech/modular-monolith-in-python/

\item \label{S36}
Hamid Reza Sharifi. Introduction to Spring Modulith. 2023. https:// www.baeldung.com/spring-modulith

\item \label{S37}
Ashley Peacock. Service Weaver: A Framework From Google For Balancing Monoliths and Microservices. 2023. https://betterprogramm-ing.pub/service-weaver-a-framework-from-google-for-balancing-m-onoliths-and-microservices-583e69b274dd

\item \label{S38}
Dmytro Lazarchuk. How Microservices And APIs Can Make Your Company Modular. 2022. https://www.forbes.com/sites/forbestechc-ouncil/2022/02/17/how-microservices-and-apis-can-make-your-co-mpany-modular/?sh=51a0185ae1cf

\item \label{S39}
Assis Zang. Creating Good Monoliths in ASP.NET Core. 2022. https:// www.telerik.com/blogs/creating-good-monoliths-aspnet-core

\item \label{S40}
Ruben Casas. Modular Monoliths: Have we come full circle? 2021. https://www.infoxicator.com/en/modular-monoliths-have-we-come-full-circle

\item \label{S41}
Adrian Ochmann. Understanding Architecture: Reasons to Build a Modular Monolith First? 2023. https://www.wearecogworks.com/ blog/understanding-architecture-reasons-to-build-a-modular-mono-lith-first/

\item \label{S42}
Lothar Schulz. Empower modular monolith – never miss these leadership advises. 2023. https://www.lotharschulz.info/2023/10/04/empo-wer-modular-monolith-never-miss-these-leadership-advises/

\item \label{S43}
Bilgin Ibryam. Distributed transaction patterns for microservices compared. 2021. https://developers.redhat.com/articles/2021/09/21/ distributed-transaction-patterns-microservices-compared

\item \label{S44}
Kevin Urrutia, Justin Trugman. Achieving Monolith Development Velocity (with a Microservice's Availability). 2023. https://www.bug-drivendevelopment.com/p/achieving-monolith-development-velocit y-with-a-microservice-s-availability

\item \label{S45}
Arkadiusz Rosloniec. Modular software architecture: advantages and disadvantages of using monolith, microservices and modular monolith. 2023. https://pretius.com/blog/modular-software-architecture/

\item \label{S46}
Modular monolith vs microservices for hybrid multi-tenancy. 2021. https://softwareengineering.stackexchange.com/questions/43236/m-odular-monolith-vs-microservices-for-hybrid-multi-tenancy

\item \label{S47}
Admir Mehanovic. Strategic Choices: Microservices vs. Monoliths for Architectural Success. 2023. https://www.penzle.com/blog/strate-gic-choices-microservices-vs-monoliths-for-architectural-success

\item \label{S48}
Aleksei Loos. Case Study: Ditching Microservices in Favor of a Modular Monolith. 2020. https://playtech.ee/blog/case-study-ditching-microservices-in-favor-of-a-modular-monolith

\item \label{S49}
Arpit Mohan. How Adopting a Modular Monolithic Architecture Enables Our OSS. 2023. https://www.appsmith.com/blog/monoliths-vs-microservices-3

\item \label{S50}
Tristan Mahinay. Modular Monolithic in Practice. 2023. https://blog. rjtmahinay.com/modular-monolithic-in-practice

\item \label{S51}
Gusto Engineering Blog. Building Toward a Modular Monolith. 2019. https://engineering.gusto.com/building-toward-a-modular-monolith/

\item \label{S52}
Spring Blog. Introducing Spring Modulith. 2022. https://spring.io/ blog/2022/10/21/introducing-spring-modulith

\item \label{S53}
Kamil Kucharski. Modular Monolith in Django. 2023. https://makimo. com/blog/modular-monolith-in-django/

\item \label{S54}
Shai Almog. Is it Time to go Back to the Monolith? 2023. https://debu-gagent.com/is-it-time-to-go-back-to-the-monolith

\item \label{S55}
Google Open Source Blog. Introducing Service Weaver: A Framework for Writing Distributed Applications. 2023. https://opensource. googleblog.com/2023/03/introducing-service-weaver-framework-for-writing-distributed-applications.html

\item \label{S56}
Service Weaver official website. https://serviceweaver.dev/

\item \label{S57}
Shiju Varghese. Monolith or Microservices, or Both: Building Modern Distributed Applications in Go with Service Weaver. 2023. https:// shijuvar.medium.com/monolith-or-microservices-or-both-building-modern-distributed-applications-in-go-with-service-a096616434fc

\item \label{S58}
Matt Campbell. Google Service Weaver Enables Coding as a Monolith and Deploying as Microservices. 2023. https://www.infoq.com/ news/2023/03/google-weaver-framework/

\item \label{S59}
Hector Valls. Service Weaver: Write monolith, deploy microservices. 2023. https://hvalls.dev/posts/service-weaver

\item \label{S60}
Preslav Rachev. Digging into Service Weaver: Dependency Injection. 2023. https://preslav.me/2023/05/12/golang-dependency-injection-in-google-service-weaver/

\item \label{S61}
Kay Ewbank. Google Introduces Service Weaver Framework. 2023. https://www.i-programmer.info/news/182-frameworks/16154-goog le-introduces-service-weaver-framework.html

\item \label{S62}
Katie Dee. Google announces Service Weaver for writing distributed applications. 2023. https://sdtimes.com/software-development/googl e-announces-service-weaver-for-writing-distributed-applications/

\item \label{S63}
Introducing Service Weaver: A Framework for Writing Distributed Applications. 2023. https://faun.dev/c/links/faun/introducing-service-weaver-a-framework-for-writing-distributed-applications/

\item \label{S64}
Google Blogs. Introducing Service Weaver: A Framework for Writing Distributed Applications. 2023. https://www.googblogs.com/introdu-cing-service-weaver-a-framework-for-writing-distributed-applicati-ons/

\end{enumerate}}

\end{document}